\documentclass[aps,prl,twocolumn,a4paper,10pt]{revtex4-1} 

\usepackage[T1]{fontenc}		
\usepackage[english]{babel}		
\usepackage[latin1]{inputenc}	
\usepackage{times}

\usepackage{amsmath}			
\usepackage{amssymb}			
\usepackage{bbm}				
\usepackage[version=3]{mhchem}		

\usepackage[colorlinks=true, a4paper=true, pdfstartview=FitV,linkcolor=blue, citecolor=blue, urlcolor=blue]{hyperref}

\usepackage{graphicx}			
\usepackage{graphics}			
\DeclareGraphicsExtensions{.pdf}
\usepackage{color}		

\newcommand{\bea}{\begin{eqnarray}}
\newcommand{\eea}{\end{eqnarray}}

\begin{document}
\title{Entropy- and flow- induced superfluid states}

\author{Johan~Carlstr\"om${}^{1}$,~and~Egor~Babaev${}^{1,2}$}
\affiliation{ 
${}^1$ Department of Theoretical Physics, The Royal Institute of Technology, Stockholm, SE-10691 Sweden \\
${}^2$ Department of Physics, University of Massachusetts Amherst, MA 01003 USA 
}

\date{\today}

\begin{abstract}
Normally the role of phase fluctuations in superfluids
 and superconductors is to drive a phase transition
to the normal state. This happens due to proliferation
of topologically nontrivial phase fluctuations in the form of vortices.
Here we discuss a class of  systems where, by contrast,
non-topological phase fluctuations can produce superfluidity.  { Here we understand superfluidity as a phenomenon which does not necessarily arises from a broken $U(1)$ symmetry,  but can be associated
with a certain class of (approximate or exact) degeneracies of the system's energy landscape giving raise to a $U(1)$-like phase.}
\end{abstract}

\maketitle

The phase transition from  superfluid to normal state is driven by phase fluctuations.
Remarkably, in the context of superfluids it was first conjectured by Onsager \cite{onsager} 
that { the the superfluid-to-normal phase transition is driven by proliferation of topological defects in the form of vortex loops}.
Due to the phase winding around a vortex, the presence of macroscopically large proliferated vortex loops
 disorders the phase $\varphi({\bf r})$ of the complex order parameter fields $\psi ({\bf r})=|\psi({\bf r})|e^{i\varphi({\bf r})}$,
leading to restoration of the $U(1)$ symmetry so that $\langle|\psi({\bf r})|e^{i\varphi({\bf r})}\rangle=0$.
This situation also takes place in $U(1)$ type-II superconductors, as 
 was established in \cite{DH}.
In that case proliferation of vortex loops restores the local $U(1)$ symmetry via inverted-XY transition (see detailed discussion in, e.g., Refs \cite{DH,peskin,sudbo,herbut}).

In two dimensions, spin-wave-like phase fluctuations play a relatively more important role: At any finite temperature
they lead to algebraic decay of correlations and thus to restoration of $U(1)$ symmetry  \cite{MW}. 
However it requires proliferation of topological defects (vortex-antivortex pairs)
to make correlations short range via the Berezinskii-Kosterlits-Thouless transition \cite{berez1,berez2,KT}.

In multi-component systems, phase fluctuations in the from of composite vortices generally results in the formation of paired states\cite{npba,nature,kuklov}. However these fluctuations still act to destroy superfluidity, albeit in some non-trivial channels. 
Indeed, more broadly we generally associate thermal fluctuations with symmetry restoration and destruction of superfluidity or superconductivity. In other physical systems indeed exceptions exists. One notable example is the Pomeranchuk and related effects\cite{Pomeranchuk,pollet}: that  \ce{^{3}He} can be liquid at zero temperature but solidifies upon heating. The transition between different lattice types in crystals (see e.g. \cite{PhysRevLett.41.702,chaikin}) is an example of a conventional phase transition between two different broken symmetries.

The question which we raise in this work is: are there situations where  fluctuations 
{\it induce} superfluidity or superconductivity rather than destroy it? 
I.e. if a situation can arise where 
 a system possesses  a channel that is not superfluid or superconducting 
in the ground state, but fluctuations cause the system to exhibit superfluidity in this channel
at, for example, elevated temperature.
We argue that indeed (quasi-)superfluid states can be generated for entropic reasons
due to phase fluctuations.

Often the superfluid phase transition
 is associated with a spontaneous breakdown of $U(1)$ symmetry, {but superfluidity can also arise without symmetry breaking} like in the aforementioned two dimensional case at finite 
temperature, or the one dimensional case at zero temperature.
Superfluidity can in principle exist in macroscopic systems that explicitly break $U(1)$ symmetry, although only weakly, thus allowing ``pseudo" Goldstone bosons.
 Here we consider a further possible generalisation 
where a superfluid mode can be associated with a $U(1)$-like degeneracy (which  indeed can be lifted by fluctuations), or nearly degeneracy
 not corresponding to a $U(1)$ symmetry of the effective potential.
 Phase fluctuations and vortices can originate from this degeneracy.
 
 In the scenario which we discuss here, a superfluid system when heated goes  
into an entropically-induced state where an additional superfluid channel is opened.
We also consider entropically induced states that are not associated with additional superfluid channels, but posses properties different from those of the ground state.

Here we discuss certain multicomponent models
because of the substantial progress 
 in creation and 
investigation of these systems recently.

In the cold atoms field, interest is driven by the extensive possibilities to realise 
multicomponent superfluid states. These systems allow tuning of various forms of inter-component interactions  with high precision. 
In the context of superconductivity, the interest 
was driven recently by discoveries of superconductors with nontrivial
multiband order parameters.

We start the discussion with 
the multicomponent Ginzburg-Landau (GL) functional, although 
our results do not rely on the existence of the GL expansion in the sense
that the effective potential does not have to be represented in this particular
form of series in powers of $\psi_i$. The free energy density is given by
\bea
F=    \sum_i  \frac{|\nabla\psi_i|^2}{2}
+\sum_{i,j}\eta_{ij}\psi_i\psi_j^*+\sum_{i,j,k,l}\nu_{ijkl}\psi_i\psi_j\psi_k^*\psi_l^*,\label{H}\;\;\;\;
\eea
where summation is conducted over $N$ components (complex fields $\psi_i$).
These components can originate from Cooper pairing in different bands  (see e.g.
 \cite{nagaosa,tesanovic,Johan2011,weston,maiti,gurevich2007}) or
Josephson-coupled superconducting layers
or correspond to different components of Bose condensed cold atoms. 
The intercomponent interaction terms in \eqref{H} 
make up the first- and second- order Josephson couplings.
Such couplings can be realised and controlled in Josephson-coupled multilayers
and Josephson-junction arrays.
Intercomponent couplings of this kind can also
be realised in cold atoms.
This problem becomes especially rich in the four component case, with 
features such as a variety of accidental degeneracies \cite{weston} appearing even at the level of only bilinear couplings.

Under certain conditions, there appear local minima of the free energy
(\ref{H}). This feature is also present  in the London limit.
Consider a three-component lattice London model containing only phases:
\bea
H =- \sum_{i,\langle j,k\rangle} \cos (\varphi^i_{j}-\varphi^i_{k})
+\sum_{k}V(\varphi^{12}_{k},\varphi^{13}_{k})\label{HXY},
\eea
where $\varphi^{ij}_{k}=\varphi^j_{k}-\varphi^i_{k}$.
The first term describes gradient energy and contains a summation over superconducting components $i$, and nearest neighbours $\langle j,k\rangle$.
The second term contains a summation over all lattice points $k$, and describes the potential energy.
 We will consider three examples of different  potentials  
 which are constructed to represent minimalistic models of the situations mentioned above, with 
 local minima in the free energy landscape.
 The potentials are chosen so that they lock phase differences, explicitly breaking the symmetry down to 
 $U(1)$. 
The local minima correspond to a different  phase locking pattern with slightly higher energy.

\begin{figure*}[!htb]
 \hbox to \linewidth{ \hss
\includegraphics[width=\linewidth]{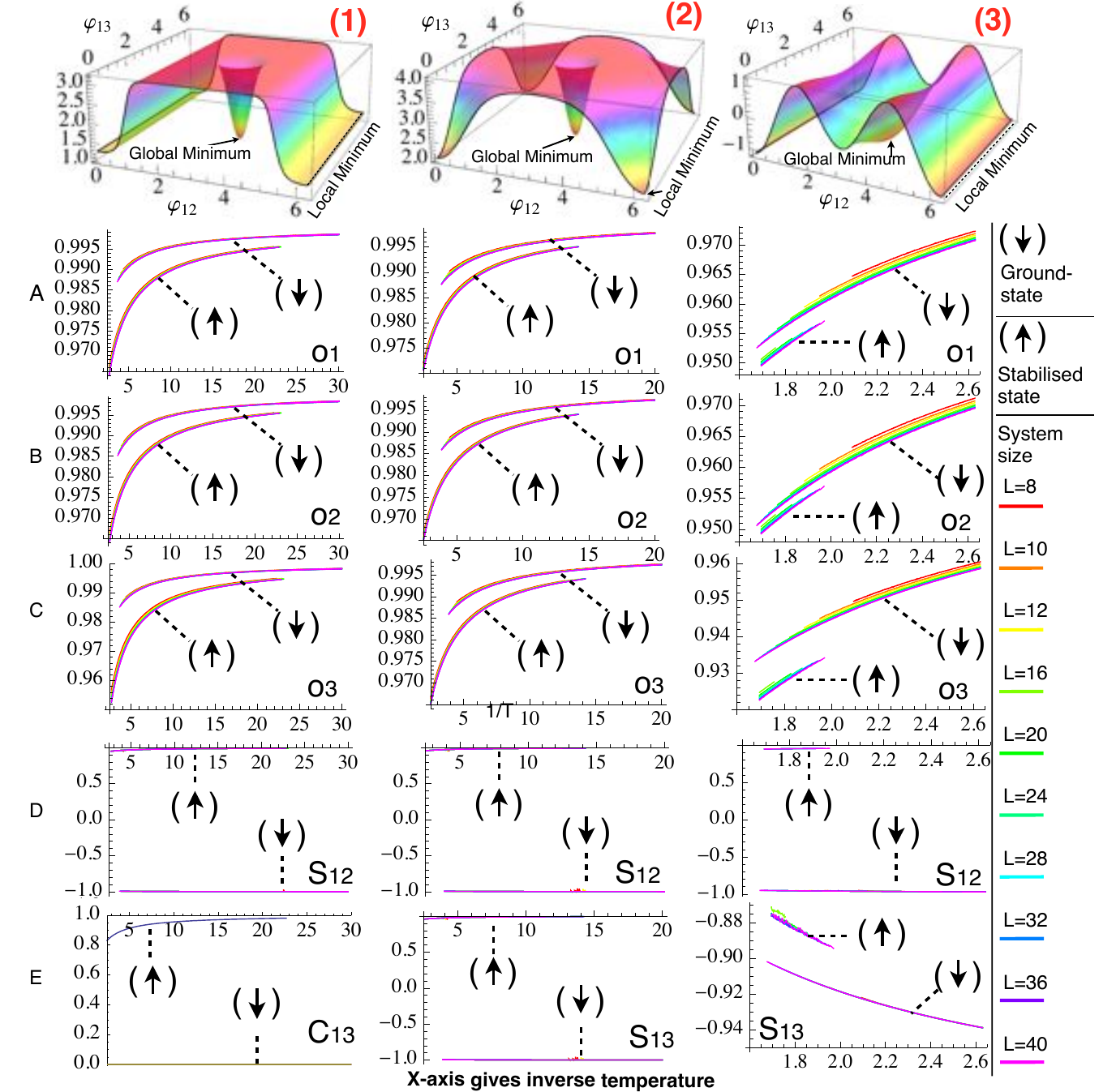}
 \hss}
\caption{
{\bf Top:} The potential {\bf (1)} exhibits a global minimum at  $\varphi_{12}=\varphi_{13}=\pi$, and a local minimum corresponding to $\varphi_{12}=0$.  The potential is given by:
$V_1(\varphi_{12},\varphi_{13})=   \tanh(3-5\cos\varphi_{12} )  
+  2.2 \tanh(8\cos\varphi_{12} + 8\cos\varphi_{13} + 16)   
$.
\\
The potential {\bf (2)} posses minima in the form of two wells: One deep but narrow situated in $\varphi_{12}=\varphi_{13}=\pi$, and a slightly more shallow, though wider located in $\varphi_{12}=\varphi_{13}=0$. The potential is given by\\
$V_2(\varphi_{12},\varphi_{13})=
2.05 \tanh(8\cos \varphi_{12}  + 8\cos \varphi_{13}  + 16) 
+ 2\tanh(2-\cos\varphi_{12}  - \cos\varphi_{13}  )$. \\
The potential {\bf (3)} posses a global minimum in $\varphi_{12}=\varphi_{13}=\pi$, and a local minimum corresponding to $\varphi_{12}=0$. The potential is given by\\
$V_3(\varphi_{12},\varphi_{13})=-\cos2 \varphi_{12} - 0.18 \cos \varphi_{12} 
+ 0.2 (1 - \cos \varphi_{12})\cos \varphi_{13} .
$
\\
{\bf Panel:} Summary of the simulation results. 
The three columns correspond to the potentials $1$, $2$ and $3$. 
 The first three rows (A-C) give the averages $o_1,\;o_2,\;o_3$ \eqref{o} for system sizes $8\le L\le 40$. These are zero if a system is in the normal state. Row (D) gives $S_{12}$ (\ref{s12}) while (E) shows the correlator $C_{13}$ (\ref{corr}) for model (1) and $S_{13}$ \eqref{s13}  for models (2,3). In (D,E) data is only displayed for the system size $L=40$. 
In all plots there are {\it two} curves corresponding to the two different states $(\downarrow\uparrow)$. These two states coexist in part of the parameter range, particularly in the models (1,2), due to hysteresis. 
 }
\label{results}
\end{figure*}

The  potentials are displayed and defined in Fig. \ref{results}. 
Two of them, $V_1$ and $V_2$, are specifically constructed as archetypical representations 
of two kinds of energy landscapes with strong features that facilitate numerical study, while
 $V_3$ is built from interaction terms of the form $\psi_i^*\psi_j$ and $\psi_i^*\psi_j^*\psi_k\psi_l$ which constitute the first and second order Josephson harmonics.
 This should give rise to two states, one that is centred around the minimum which we denote $(\downarrow)$, and another state centred around the local minimum which we denote ($\uparrow$). 
Another feature of these potentials is that they exhibit a comparatively steep slope around the global minimum, whilst the local minimum is surrounded by a flatter slope in the  energy landscape. 
Consequently, the energy cost of phase-difference excitations is smaller in the state ($\uparrow$) than in ($\downarrow$). 

We are interested in the role of phase fluctuations in these systems. The fact that these are energetically cheaper in the state ($\uparrow$) implies a lower free energy at a certain temperature, resulting in a transition to a an entropically stabilised state. 
The normal role of phase fluctuations is to restore the $U(1)$ symmetry, but in the models (1,3) the situation is reversed as the state $(\uparrow)$ actually posses an additional superfluid mode.

To investigate the entropically stabilised superfluid state we have conducted  Monte Carlo simulations based on the metropolis algorithm with two types of update:
The first is a local update concerning only a local spin. 
The second is a global update where one of the phases is selected and an update is proposed so that $\varphi_i\to\varphi_i+\Delta\pi$ on all lattice points simultaneously. See also remark \cite{MonteCarlo}. 

During the simulation several quantities were monitored: The coherence of the individual phases was measured as
\bea
o_\gamma(T)=\Big\langle \Big| \frac{1}{L^3}\sum_{j}e^{i\varphi_{\gamma,j}}\Big|  \Big\rangle \label{o}
\eea 
where index $1\le \gamma\le 3$ denotes the different phases. 
To identify the two states we also introduce 
\bea
S_{12}=\langle \cos(\varphi_2-\varphi_1) \rangle \label{s12}
\eea
In the state ($\downarrow$) we should have $S_{12}\sim-1$ while in the state ($\uparrow$) we expect $S_{12}\sim1$. 
To determine the properties of the sector $\varphi_{1,3}$, we introduce the following two functions: 
 When the system is in a state where $\varphi_{13}$ has
an appreciable  mass (which turns out to be the case in model 2 but in fact also 3) we calculate
\bea
S_{13}=\langle \cos(\varphi_3-\varphi_1) \rangle \label{s13}
\eea
which has a preferential value that reflects degeneracy lifting. 
When the mode becomes superfluid (which occurs in model 1), this is best illustrated by the correlator
\bea\nonumber
C_{13}=\frac{1}{N} \int dr \Big[\big\langle X(r)\cdot X(0)  \big\rangle
-  \big\langle  X(r)\big\rangle\cdot \big\langle X(0)  \big\rangle\Big],\\
X=\{\cos \varphi_{13},\sin\varphi_{13}\},\;\;N=L^{D}. \label{corr}
\eea
Here superfluid modes are recognised as having non-zero correlations even in large systems. 
Consequently, in a large system $C_{13}$ takes a finite value in a superfluid while it approaches zero for a massive mode. 

The results of the simulations are summarised in Fig. \ref{results}. All three models exhibit a transition to an entropically stabilised   state that is 
driven by non-topological phase fluctuations.
These fluctuations are energetically cheaper in the state ($\uparrow$), something that is reflected in the fact that the phase coherence ($o_i$)  at a given coupling strength is smaller in this state (A-C).  
The  transition is associated with a change of the parameter $S_{12}$, which is $\sim-1$ in the state ($\downarrow$) and $\sim+1$ in ($\uparrow$) (D).

 The simplest situation
takes place in the model (2). It has two  minima: a deep and narrow well ($\downarrow$) and a slightly more shallow well with a flatter slope ($\uparrow$) with no additional degeneracy. Thus, in this case the phase fluctuations can drive a transition between two states with different phase lockings and different normal modes, giving raise to substantially different coherence lengths. 

In the model (1), the ground state ($\downarrow$) breaks $U(1)$ symmetry, while in the state ($\uparrow$), the $U(1)$-like degeneracy is broken in addition.
This is reflected in the appearance of a nonzero correlator  ($C_{13}$) and thus entropically induced superfluidity

The model (3) is similar to (1) in the sense that it exhibits a ground state where the energy is invariant under $U(1)$ transformations 
as well as a local minimum corresponding to $\varphi_{12}=0$. In the local minimum the energy is independent of $\varphi_{13}$, implying that it has an additional degeneracy. However, in this state the energy is only independent of $\varphi_{13}$ when $\varphi_{12}=0$ exactly, i.e. precisely at the bottom of the 'valley'. At any nonzero temperature, thermal fluctuations renders this mode massive (although with a comparatively very small mass). This is reflected in the fact that $S_{13}$ (shown in E3) is nonzero even for the state ($\uparrow$). At finite length scale, 
this state however shares  properties with a $U(1)\times U(1)$ superfluid.

Importantly, in  these type of systems, the transitions  between the states can occur not only because of thermal fluctuations, but also as a result of an applied phase twist.  
This can be demonstrated as follows: to simplify calculations, consider a system similar to model (1), but with an infinitely narrow well. Let the energy be $0$ in the well, $V$ in the valley and $U$ everywhere else so that $0< V<U$.
Suppose $\varphi_{1,3}$ is twisted by $N2\pi$ along a system of size $L$. In the state ($\downarrow$) the twist results in kinks of width $l_w=\pi\sqrt{2/U}$ and energy $E_w=2\pi\sqrt{2U}$. In the state $(\uparrow)$ the twist instead results in an evenly distributed phase gradient. The energy difference between the two solutions can then be written
\bea
E_\uparrow-E_\downarrow = L\Big( -\rho E_w + 2\pi^2\rho^2 +  V \Big),\;\rho=\frac{N}{L}.
\eea
This expression is valid provided $0\le \rho\le 1/l_w$ (since the size of the kinks is $l_w$).
If $\rho=0$, this is positive and $(\downarrow)$ is the lowest energy state. Inserting $\rho= 1/l_w$
 we instead obtain $E_\uparrow-E_\downarrow =V-U<0$, and the lowest energy state is $(\uparrow)$. 
Thus, above some critical phase twist $\rho_c$, it is energetically beneficial to go to the superfluid state ($\uparrow$)
where the system can take advantage of the (near) degeneracy in the energy landscape.
 Below it, ($\downarrow$) becomes the preferred state. 

Finally, we adress the question of mass generation through thermal fluctuations. 

The fact that the induced superfluid state does not correspond to an underlying symmetry of the Hamiltonian means that fluctuations can result in this mode becoming massive. 

\begin{figure}[!htb]
\includegraphics[width=\linewidth]{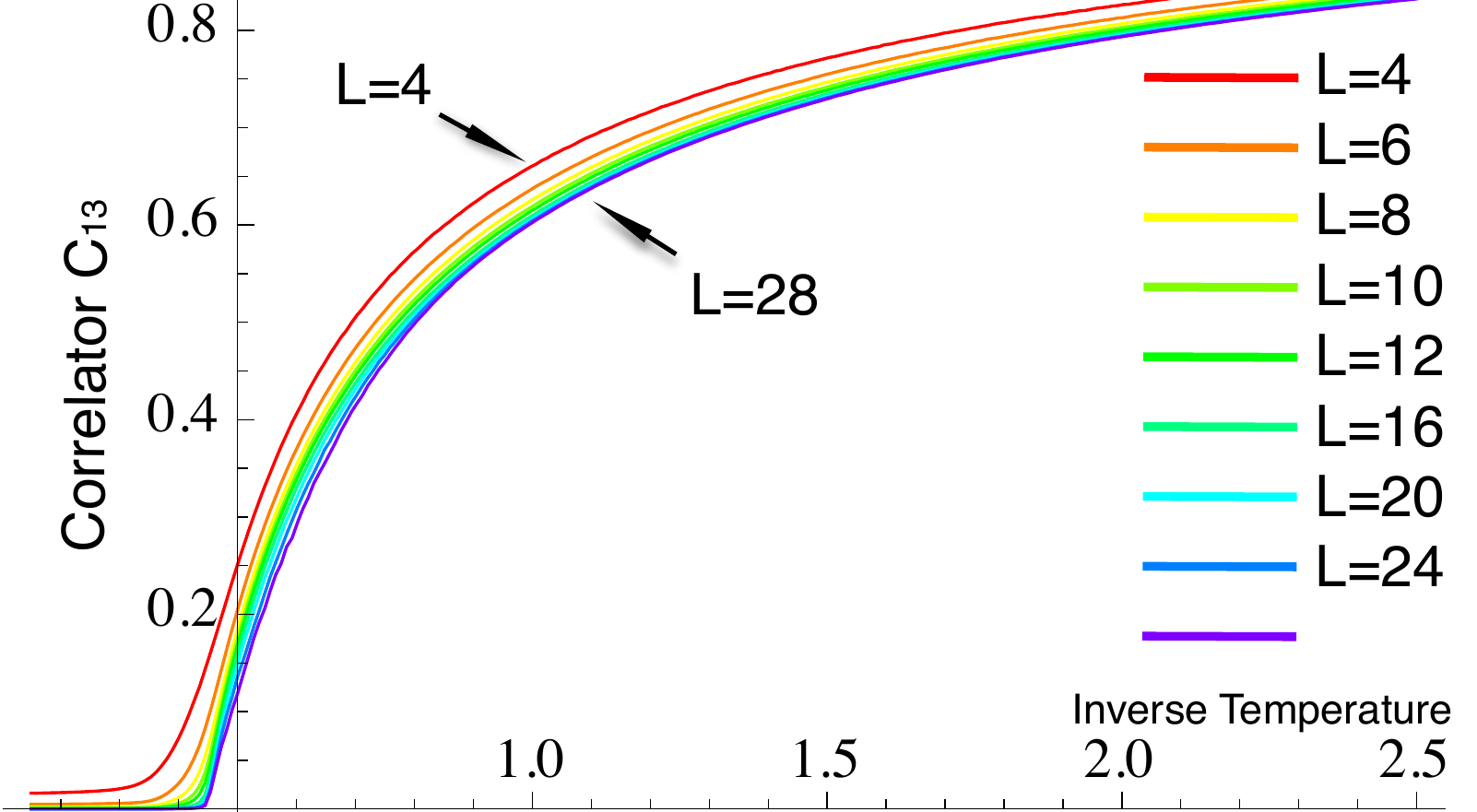}
\caption{
Correlator $C_{13}$ for model (1) at higher temperature. System sizes range from $4$ to $28$. The point on the horizontal axis $\beta\approx 0.45$, where $C_{13}$ approaches zero corresponds to the  transition where superfluidity is destroyed by vortex proliferation. 
}
\label{High temperature correlator}
\end{figure}

The extent to which this occurs very much depends on the shape of the potential, as well as the temperature. 
In the model (3) this phenomena is directly visible even at relatively low temperature in a finite system, where it renders $S_{13}$ nonzero in the state ($\uparrow$). 
In contrast, the model (1) is unaffected by this process   within the accuracy of the simulations.
When the fluctuations render the mode massive the system behaves as a superfluid on length scales smaller than the inverse mass of the mode. 
The Fig. \ref{High temperature correlator} shows the correlator $C_{13}$ at higher temperature for model (1) and system sizes $4\le L\le 28$. At these sizes the $\varphi_{13}-$sector remains superfluid until vortex proliferation takes place at $\beta\approx 0.45$. 

{ In conclusion we demonstrated the existence of the class of systems
that posses a channel that is not superfluid in the ground state but upon heating becomes superfluid   due to entropy associated with non-topological
phase fluctuations. 
 We also discussed that such systems can  have upper and lower critical superfluid velocities. The superfluid response in the considered example is associated with a (near) degeneracy which does not originate from a symmetry, yet it allows to define a superfluid phase variable and quasi-topological defects (vortices). It would be
 especially interesting to realise such systems in cold atoms experiments.}

\begin{acknowledgments}
We thank A. Kuklov, B. Svistunov, N. Prokof'ev, M. Wallin and J. Lidmar for discussions.
This work was  supported  by the Knut and Alice Wallenberg Foundation through a
Royal Swedish Academy of Sciences Fellowship, by the Swedish Research Council, and by the US National Science Foundation under the CAREER Award DMR-0955902.
The computations were performed using resources provided by the Swedish National Infrastructure for Computing (SNIC) at the National Supercomputer Center in Linkoping, Sweden. 
\end{acknowledgments}


\begin{thebibliography}{23}
\bibitem [{Mon()}]{MonteCarlo}%
  \BibitemOpen
  \href@noop {} {}\bibinfo {note} {The updates are accepted with probability
  $e^{-\beta \Delta E}$. For each model, 256 different coupling strengths
  (inverse temperatures) were selected. For each coupling strength, two
  simulations corresponding to the two possible states
  ($\uparrow$,$\downarrow$) were prepared. The equilibration was conducted in
  two stages:First, an initial 50,000 local updates were carried out (i.e. each
  lattice point was subject to 50,000 proposed updates).Second, another 50,000
  updates were carried out. But this time, every 3 updates, a global update was
  proposed.After that, the equilibration, data was collected as follows:Local
  updates were proposed 200 times for every lattice point. For every 3 such
  updates, a global update was proposed. Then, simulation data was
  collected.This process was repeated 28,000-100,000 times depending on system
  size, resulting in the same number of data points. The simulations were
  conducted on a cubic lattice with periodic boundary conditions and system
  sizes $ 8\le L\le 40$. Finally, points in the parameter space where the
  system keeps tunnelling between the two states were discarded (this mainly
  occurred for smaller systems in model 3).}\BibitemShut {Stop}%
\end{thebibliography}
\end{document}